\documentclass[10pt,twocolumn,floatfix,prb,superscriptaddress]{revtex4-1}
\usepackage[dvipsnames]{xcolor}
\usepackage{amsmath,amssymb,amsthm,mathrsfs,amsfonts,dsfont,amstext} 
\usepackage{textcomp,pbox}
\usepackage[export]{adjustbox}
\usepackage{bm}
\usepackage{dcolumn,booktabs,url}
\usepackage[scaled]{helvet}
\usepackage{sansmath,gensymb}
\usepackage{tikz,graphicx,transparent}
\usepackage{multirow}
\usepackage[separate-uncertainty = true]{siunitx}
\usepackage{comment}
\usepackage{physics}
\usepackage{pdfpages}
\usepackage{array}

\makeatletter
\AtBeginDocument{\let\LS@rot\@undefined}
\makeatother

\newcolumntype{C}[1]{>{\centering\let\newline\\\arraybackslash\hspace{0pt}}m{#1}}

\usepackage[colorlinks=true]{hyperref}
\usepackage{graphicx}

\hypersetup{
     colorlinks   = true,
     citecolor    = red,
     linkcolor    = blue,
     urlcolor     = red     
}

\graphicspath{{./figs/}}

\begin{document}
% no post selection. Direct measurement of phase shift induced by a quantum emitter.
% Giant non linear phase shift.
% Switchin the phase of a light beam using a single quantum emitter?
\newcommand{\TitleName}{Direct observation of non-linear optical phase shift induced by a single quantum emitter in a waveguide}
\title{\TitleName}

\newcommand{\AffCPH}{Center for Hybrid Quantum Networks (Hy-Q), Niels Bohr Institute, University~of~Copenhagen, DK-2100 Copenhagen~{\O}, Denmark}
\newcommand{\AffBasel}{Department of Physics, University of Basel, Klingelbergstra\ss e 82, CH-4056 Basel, Switzerland}
\newcommand{\AffBochum}{Lehrstuhl f\"ur Angewandte Festk\"orperphysik, Ruhr-Universit\"at Bochum, Universit\"atsstra\ss e 150, 44801 Bochum, Germany}

%% Author order to be changed! just a list of people!
\author{Mathias J.R. Staunstrup}
\affiliation{\AffCPH{}}

\author{Alexey Tiranov}
\affiliation{\AffCPH{}}

\author{Ying Wang}
\affiliation{\AffCPH{}}
%\author{Vassiliki Angelopoulou}
%\affiliation{\AffCPH{}}

\author{Sven Scholz}
\affiliation{\AffBochum{}}

\author{Andreas D. Wieck}
\affiliation{\AffBochum{}}

\author{Arne Ludwig}
\affiliation{\AffBochum{}}
\author{Leonardo Midolo}
\affiliation{\AffCPH{}}
\author{Nir Rotenberg}
\affiliation{\AffCPH{}}

\author{Peter Lodahl}
\affiliation{\AffCPH{}}
\email{lodahl@nbi.ku.dk}
\author{Hanna Le Jeannic}
\affiliation{\AffCPH{}}
\email{hanna.le-jeannic@cnrs.fr}

\date{\today}

\begin{abstract}
Realizing a sensitive photon-number-dependent phase shift on a light beam is required both in classical and quantum photonics. It may lead to new applications for classical and quantum photonics machine learning or pave the way for realizing photon-photon gate operations. Non-linear phase-shifts require efficient light-matter interaction, and recently quantum dots coupled to nanophotonic devices have enabled near-deterministic single-photon coupling. We experimentally realize an optical phase shift of $0.19 \pi \pm 0.03$ radians ($\approx 34$ degrees) using a weak coherent state interacting with a single quantum dot in a planar nanophotonic waveguide. The phase shift is probed by  interferometric measurements of the light scattered from the quantum dot in the waveguide. The nonlinear process is sensitive at the single-photon level and can be made compatible with scalable photonic integrated circuitry. The work may open new prospects for realizing high-efficiency optical switching or be applied for proof-of-concept quantum machine learning or quantum simulation demonstrations. 
\end{abstract}

\maketitle

Optical nonlinearities are at the core of many modern applications in photonics. If sensitive at the level of single light quanta, they may be applied to realize fundamental quantum gate operations for photonic quantum computing or advanced quantum network implementations \cite{Chang2014,Uppu2021}.
The nanophotonics platform could potentially be scaled up to realize large-scale nonlinear quantum photonic circuits, as required, e.g., in quantum neural networks\cite{Steinbrecher2019}. Strong optical nonlinearities can be achieved using single emitters such as molecules or quantum dots (QDs) embedded in photonic waveguides or cavities\cite{Lodahl2015,Turschmann2019} due to the tight confinement of light to reach light-matter coupling efficiencies near unity \cite{Arcari2014}.
In the waveguide geometry, a narrow-band single-photon wavepacket is deterministically reflected upon resonant interaction with a highly coherent two-level quantum emitter, while two-photon wavepackets are partly transmitted due to the saturation of the emitter \cite{Shen2007,LeJeannic2022}, allowing for realizing deterministic quantum operations such as photon sorters \cite{Witthaut2012,Yang2022}.
In most experiments and protocols, however, the focus has been on measuring the intensity modification of a light field after interaction with the emitter\cite{Javadi2015,Faez2014Nov,Antoniadis2022,Pscherer2021Sep}, either in transmission ($I_{t}$) or in reflection ($I_{r}$).
However, the direct measurement of the essential phase response of the nonlinear interaction requires interferometric measurement of the optical response of the quantum emitter.

Emitter-induced phase shifts were demonstrated in atomic ensembles, either at room temperature or in magneto-optical traps, \cite{Zibrov1996May}, and using  trapped single atoms or ions \cite{Aljunid2009,Fischer2017Jan}.
However, there, the relatively weak light confinement achievable by tightly focusing a free-space laser beam, limited the achievable phase shift from a single atom to a few degrees\cite{Aljunid2009}. Free-space, high finesse cavities were considered to increase the light-atom coupling \cite{Turchette1995,Reiserer2013Dec}, as well as their nanophotonic equivalents \cite{Volz2011Jul,Volz2014,Tiecke2014Apr}, enabling to drastically increase the reachable phase shift by single atoms, although at the cost of greater experimental complexity.
In parallel, solid-state emitters have been considered a promising platform due to their ease of integration with nanophotonic structures \cite{Fushman2008May} and significant phase shifts have been demonstrated in nanocavities \cite{Wang2019}. In nanophotonic waveguides, such effects have remained limited to a few degrees, hindered by the light-emitter coupling efficiency \cite{Pototschnig2011Aug}.
Among them, single QDs embedded in photonic waveguides can potentially reach very pronounced single-photon phase shifts, thanks to the high single-mode coupling efficiency \cite{Arcari2014} and nearly lifetime-limited emission lines \cite{Pedersen2020}.

\begin{figure}
 \includegraphics[width=0.9\linewidth]{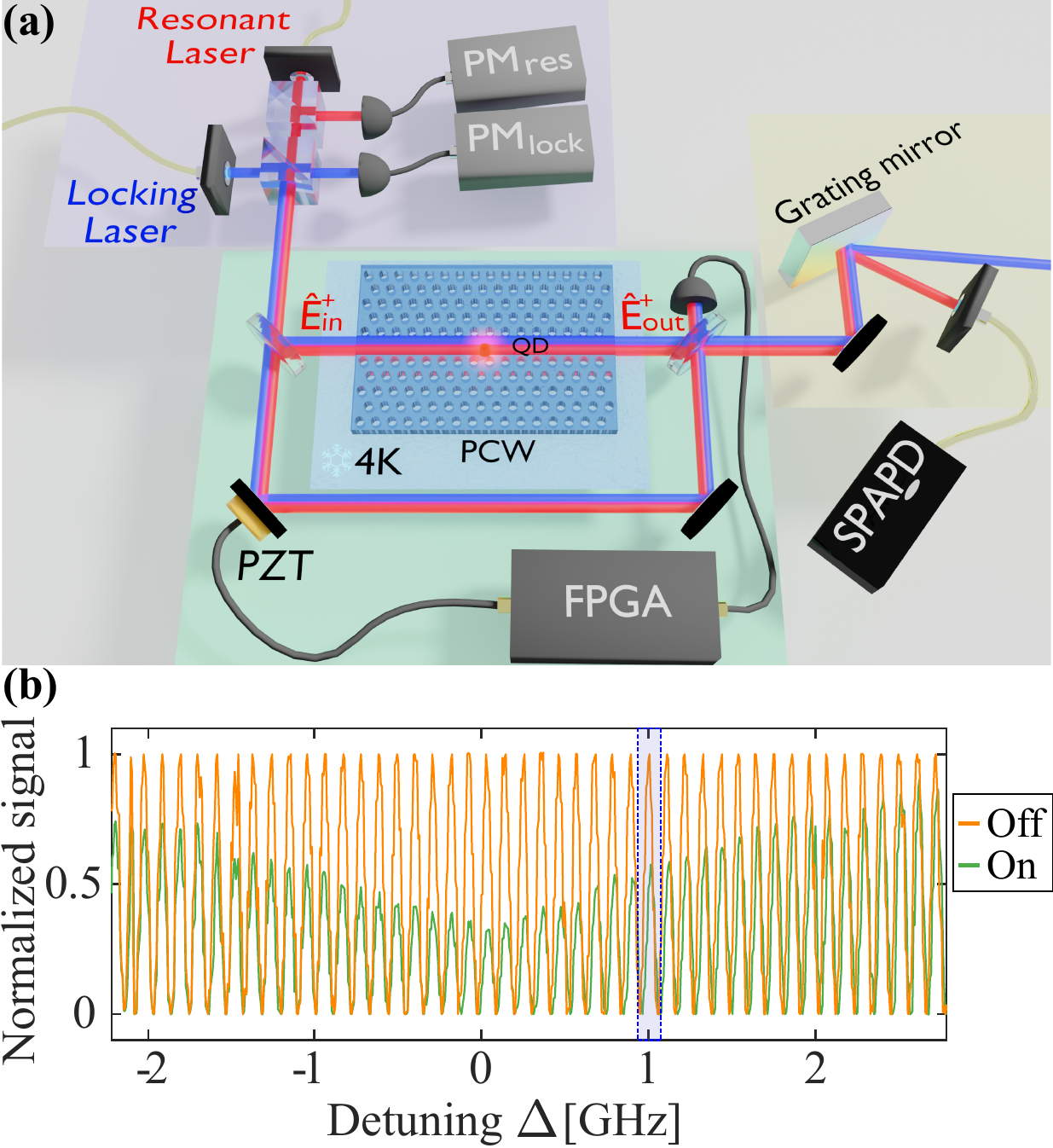}
 \caption{(a) Experimental setup: a Mach-Zehnder interferometer is used to measure the phase shift caused by a single quantum dot (QD) in a photonic crystal waveguide(PCW) cooled to 4~K. The interferometer is locked using a two-color scheme, where a far-detuned laser (blue) is used as a reference, and a feedback loop is implemented with a FPGA and a piezo-electric transducer (PZT). The low-power, resonant interference signal (red) is separated from the higher-power locking beam (blue) through a grating mirror. The filtered signal is then captured by a single-photon avalanche photodiode (SPAPD). $\text{PM}_\text{{res}}$ and $\text{PM}_\text{{lock}}$ are the two power meters used to stabilize the laser powers.
(b) Evolution of the interference signal with detuning of the resonant laser (relative to the most pronounced QD transition) when the QD is tuned \emph{on} (green). Same laser tuning range interference evolution when the QD is switched \emph{off} (orange) through the application of an electric field across the QD (DC-Stark effect). A zoom-in of the blue area is presented in Fig.~\ref{fig:figure2} (b).}
 \label{fig:figure1} 
\end{figure}

In a waveguide, the transmission coefficient is defined as $t=\frac{\langle \bf{\hat{E}}_{out}\rangle_{ss}}{\langle \bf\bf{\hat{E}}_{in}\rangle_{ss}}$, where $\bf{\hat{E}}_{in}$ and $\bf{\hat{E}}_{out}$ are the input and output field operators, respectively (see Fig.~\ref{fig:figure1}(a)), evaluated in the steady state (ss). 
The phase shift is expressed as its argument $\phi= arg(t)$. 
In the case of a lifetime-limited quantum emitter of decay rate $\gamma$ and bidirectional (isotropic) interaction, the maximum single-photon phase shift achievable on resonance reaches $\pi/2$, in the limit where the light-matter coupling efficiency (the $\beta$-factor) reaches unity \cite{Lodahl2015}.
For $\beta \neq 1$, the phase shift is maximum for a light-emitter detuning of $\Delta=\pm \gamma  \frac{\sqrt{1-\beta}}{2}$ 
%\begin{equation}
%\begin{split}
 %|\phi|_{max} &=\textrm{Max}_{\Delta}(arg(t))\\
 %&=tan^{-1}\left(\frac{\beta(2-i\sqrt{1-\beta})-2}{\beta-2}\right)
 %\end{split}
%\end{equation}

\begin{equation}
 |\phi|_{max} =tan^{-1}\left(\frac{\beta }{2 \sqrt{1-\beta }}\right)
\end{equation}

(see Supplementary Information for the detailed calculation of the transmission coefficient). 
Recently, a photon-scattering reconstruction method was implemented to indirectly infer a phase shift of $~ 0.22\pi$ \cite{LeJeannic2021}. Here, we demonstrate the direct measurement of a single-photon phase shift induced from the interaction with a QD in a nanophotonic waveguide by implementing interferometric measurements. 

The measurement setup, sketched in Fig.~\ref{fig:figure1}(a), consists of an approximately 3m long Mach-Zehnder interferometer built on top of a closed-cycle cryostat, where the nanophotonic chip is cooled down to 4K.
A continuous-wave laser is sent to one of the interferometer arms containing a GaAs photonic crystal waveguide with an InGaAs QD embedded inside (for more details on the sample fabrication, see \cite{Kirsanske2017Oct}). After interaction with the QD, the signal is coupled out of the waveguide chip and interfered with the reference arm (the local oscillator, LO). The achieved interferometer visibility is $v\approx0.65$, mainly limited by the imperfect mode matching between the LO and the light out-coupled from the chip's gratings. The limited visibility only affects the signal-to-noise ratio of the measurement but suffices for resolving the narrow spectral features of the QD resonances. The resulting interference signal is then sent to a single-photon detector.
To stabilize such a long interferometer, which is sensitive to sub-wavelength-scale vibrations, we apply a second laser, the locking laser, to measure and implement fast feedback corrections on the optical path. This laser is blue-detuned by 7.5 nm from the QD transition at 941 nm to avoid any interaction with the emitter, and at a much higher power than the few-photon resonant laser. The interferometer is locked using an FPGA (Field Programmable Array, Red Pitaya) programmed to act like a lock-in amplifier followed by a proportional–integral–derivative controller\cite{Luda2019_RPPID}. The feedback signal is sent to a piezoelectric transducer (PZT) mounted mirror to compensate for any change in phase not originating from the quantum emitter (see Supplemental Information for more details on the experimental setup). Finally, the locking laser is filtered from the signal using a grating filter setup.

\begin{figure}
 \includegraphics[width=0.9\linewidth]{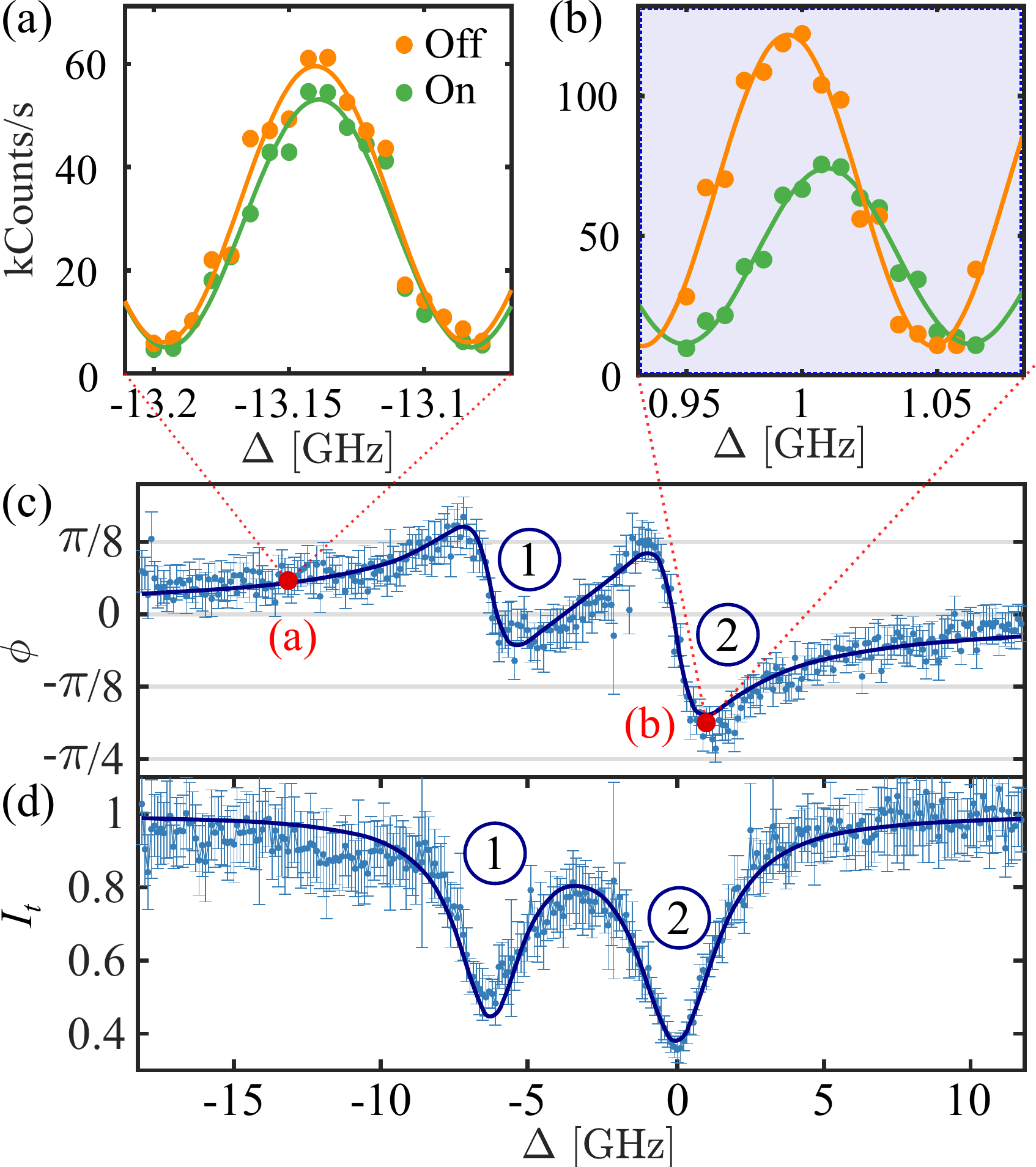}
 \caption{(a) and (b) Direct interferometric data with the emitter tuned \emph{on} (green) and \emph{off} (orange) resonance using the external electric field, for two different laser-emitter detunings. The measurement points are plotted along with corresponding sinusoidal fits (solid line). The data in (b) correspond to the detuning area marked in blue in Fig. \ref{fig:figure1}(b). (c) and (d) Extracted respective phase shift and transmission for the two dipoles, labeled 1 and 2. The solid lines correspond to the fit of the data to the theory.}
 \label{fig:figure2}
\end{figure}

To probe the phase shift, the frequency of the resonant laser is swept across the QD resonance to measure the resulting interference signal, while the locking laser frequency stays fixed. 
 We tune the resonance frequency of the QD with a voltage applied across the sample by virtue of the DC-Stark effect \cite{Kirsanske2017Oct}, allowing us to compare the on- and off-resonance cases, respectively (See Figure \ref{fig:figure1}(b)), and determine directly and accurately the phase shift induced by the QD.
Figure \ref{fig:figure2}(a) and (b) presents two examples of signals at different laser detunings. Away from resonance (Figure \ref{fig:figure2}(a)), no significant intensity and phase change are observed, while near resonance (Figure \ref{fig:figure2}(b)), the fringe contrast and phase changes when the QD is set to be resonant with the laser field.
Through a single measurement, we can thus infer both the phase and intensity changes experienced by the light field due to the interaction with the QD.
The results are presented in Fig. \ref{fig:figure2}, where the phase (c) and intensity (d) spectra of the two dipole transitions of the QD neutral exciton, labeled (1) and (2), are displayed.
We fit the phase and intensity data of both dipoles simultaneously (See Supplementary Information), and infer the maximal phase shifts to be $\phi_{max,1}=(-0.06\pm0.03) \pi$ and $\phi_{max,2}=(-0.19 \pm0.03) \pi$ radians, respectively. 
The observed phase shift is about thirty times larger than previously demonstrated using a direct Mach-Zehnder interferometric measurement \cite{Aljunid2009}. In contrast, using a heterodyne technique, phase shifts induced by organic molecules has shown up to $0.017\pi$ \cite{Pototschnig2011Aug} and $0.37\pi$ \cite{Wang2019} radians, the later being cavity-embedded. Further experiments show a $ \simeq \pi$ phase shift, in the reflection off an atom coupled to a cavity\cite{Tiecke2014Apr,Jechow2013}. Additionally, our experiment presents a method of directly measuring the total transmission response across the resonance of an emitter in a waveguide, which is only limited by the recorded count rate. Fig. \ref{fig:figure2}(b)  shows the experimental  data  for a 100ms integration time.  

%\cite{Pototschnig2011Aug} : molecule,in waveguide Vahid group, indirect measurement
%\cite{Wang2019}molecule,in cavity, Vahid group, indirect measurement
%\cite{Tiecke2014Apr}: pi phase shift Lukin group 1atom 1 nanocavity, self interference in reflection. 
%\cite{Volz2014}: pi phase shift Rauschenbeutel group 1atom 1 resonator, in reflection. 

%\cite{Jechow2013} Ion free space, self interferfening, 1 radian
%\cite{Tiarks2016Apr}: Rempe, cloud of Rydberg in transmission.

\begin{figure}
 \includegraphics[width=0.9\linewidth]{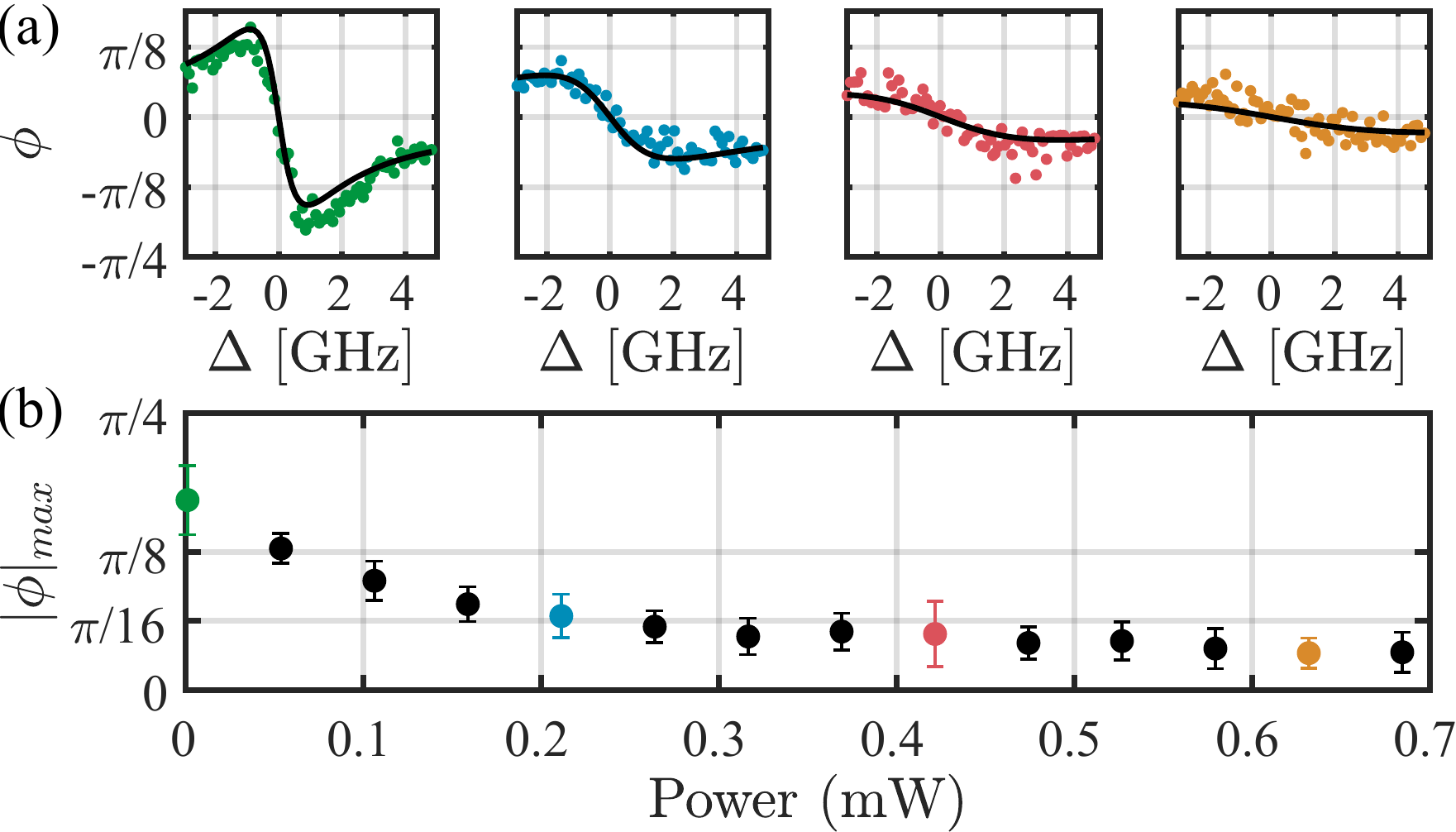}
 \caption{(a) Measurements of the phase response of the QD versus detuning and for different excitation powers. The solid lines are the fit to the theory of the overall data set. (b) Maximum measured experimental phase shift as a function of input power (measured at $\text{PM}_\text{{res}}$, see Fig. \ref{fig:figure1}(a). The colored points correspond to the data shown in (a)}
 \label{fig:figure3}
\end{figure}

Next, we examine the saturation of the phase shift in order to investigate its nonlinear response to changes in the incoming laser power. We consider dipole transition (2).
In Fig.~\ref{fig:figure3}(a), we show several spectra taken at different laser power levels and the corresponding fitting of the full saturation behavior (see Supplementary Informaion), which is fully consistent with the data presented.
For each power level, we determine the maximum experimentally observed phase shift and investigate the nonlinear behavior as the QD saturates, see Fig.~\ref{fig:figure3}(b).  By using the experimental parameters extracted previously, we estimate that the saturation happens at a mean photon flux of $n_c\sim0.33$ photons interacting with the QD during its lifetime (See \cite{LeJeannic2021} and Supplementary Information), well below the single-photon level.

\begin{figure}
 \includegraphics[width=0.9\linewidth]{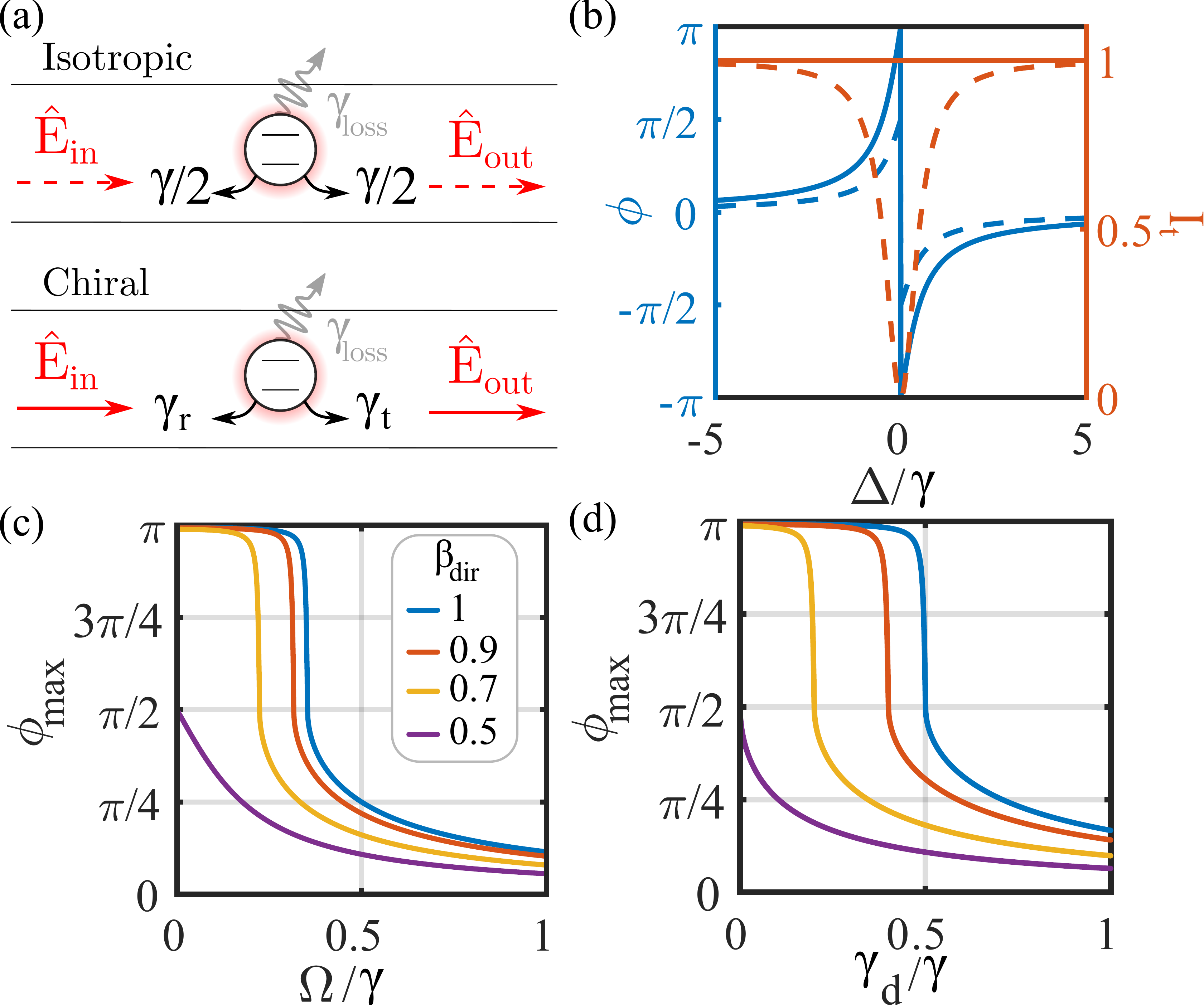}
 \caption{(a) Illustration of the scattering configuration for an isotropic (top) and a chiral (bottom) waveguide. In the latter, the reflection and transmission decay rates ($\gamma_r$ and $\gamma_t$ respectively) differ (b) Phase and transmission response for isotropic (dashed) and chiral (full line) cases for an ideally coupled system (no dephasing and no lossy channel). (c) Maximal phase shift $\phi_{\text{max}}$ as a function of the driving Rabi frequency $\Omega$ for different directional coupling efficiencies $\beta_{dir}=[1$ (blue), $0.9$ (red) $0.7$ (yellow) $0.5$ (purple)$]$. $\beta_{dir}=0.5$ corresponds to the case of an isotropic waveguide with $\beta=1$. (d) $\phi_{\text{max}}$ as a function of the pure dephasing rate $\gamma_d$ for a series of coupling efficiency $\beta_{dir}=$\{1 (blue), 0.9 (red) 0.7 (yellow) 0.5 (purple)\}}.
 \label{fig:figure4}
\end{figure}

The experimentally extracted phase shifts are limited by the coupling efficiency and decoherence of the QD and future experiments on fully lifetime-limited QD transitions \cite{Pedersen2020} should allow observing a phase shift approaching $\pi/2$. Going beyond this would even be possible in the setting of chiral quantum optics \cite{Lodahl2017Jan} where directional coupling entails that the reflective "loss channel" can be strongly suppressed Fig. \ref{fig:figure4}(a) schematically illustrates the isotropic and chiral cases, respectively.
In the ideal chiral case, the maximum possible phase shift of $\pi$ can be realized, the ultimate goal for quantum phase gates\cite{Ralph2015, Chang2014, Borregaard2019}. In contrast, the transmitted intensity would be unchanged at resonance, see Fig.~\ref{fig:figure4} (b), i.e. no photons are lost and the scattering is thereby deterministic in transmission. Such a single-photon response, however, would be undetectable in intensity measurements and therefore require the interferometric method demonstrated here. 
It is interesting to further exploit the unusual behavior of the phase response in the chiral geometry. When the input light intensity is increased, a very abrupt phase response is predicted (see Fig.~\ref{fig:figure4}(c)), unlike in the symmetric configuration. Indeed, towards saturation the transmission coefficient at resonance (which is real) changes from a negative value to a positive value, resulting in a sudden shift in the phase from $\pi$ to 0. This may find applications as an all-optical phase-switch \cite{Chen2013,Pototschnig2011Aug,Tiecke2014Apr}. Similarly a sharp transition can be found while varying the dephasing rate (see Fig.~\ref{fig:figure4}(d)), which means it may be applicable as  an ultra-sensitive probe of environmental decoherence processes of the QD.
Finally, we rediscover that the case of ideal directionality is equivalent to an ideal emitter in an isotropic waveguide when the efficiency decreases by half due to saturation ($\Omega \geq\frac{\gamma}{2\sqrt{2}}$), dephasing ($\gamma_{dp} \geq \gamma/2$), or coupling inefficiency ($\beta_{dir}\leq 1/2$).

In summary, we have developed an interferometric method for measuring the nonlinear phase shift of light caused by a single quantum emitter and used it to obtain a record-high phase response in a waveguide. 
These results may open up for a wide range of applications on how to realize deterministic quantum phase gates in photonic circuits\cite{Ralph2015,Chen2021} as a basis for  quantum non-demolition measurements \cite{Reiserer2013Dec,Volz2011Jul} or deterministic generation of optical Schrödinger cat states\cite{Wang2005Aug}, when combined with accurate spin control\cite{Borregaard2019,Duan2004,Tiarks2016Apr}. 
Additionally, the quantum emitter phase shift may be  applied as the quantum nonlinear operation required in quantum optical neural networks\cite{Steinbrecher2019}, where even moderate nonlinear phase shifts have been shown to suffice for improving the implementation of Bell-state detectors\cite{Pick2021,Ewaniuk2023}. 
Chiral light-matter interaction promises to improve the phase response even further, and in such a configuration interferometric measurements are required in order to detect the single-photon scattering processes, and the complex phase response acquired by optical pulses constitutes an interesting future direction of research that also may shed new light on applications of the emitter nonlinearity.

\section{Acknowledgments}
We thank Vasiliki Angelopoulou for her help at the early stages of the experiment.
We acknowledge funding
from the Danish National Research Foundation (Center of
Excellence “Hy-Q,” Grant No. DNRF139) and from
the European Union’s Horizon 2020 research and innovation
programmes under Grant Agreements No. 824140 (TOCHA,
H2020-FETPROACT-01-2018). This project has also received funding from BMBF 16KIQS009.

\bibliography{biblio}

\newpage
\section{Supplementary Information}

\subsection{Transmission of the emitter-waveguide system}

The quantum dot is modeled as a two-level system (TLS) with ground-and excited states $\ket{g}$ and $\ket{e}$. 
The Hamiltonian describing the light-emitter interaction can be written as\cite{Turschmann2019}:
\begin{equation}
   \hat{H}=-\hbar \Delta \hat{\sigma}_{eg}\hat{\sigma}_{ge}+\hbar\omega_p{\bf\hat{f}^\dagger({\bf r})}{\bf\hat{f}({\bf r})}-{\bf\hat{d}}\cdot{\bf\hat{E}({\bf r})}
\end{equation}
The first term describes the emitter dynamics with $\Delta=\omega-\omega_{TLS} $ as the detuning between the driving field of frequency $\omega$ and the two-level system resonance $\omega_{TLS}$. $\hat{\sigma}_{ij}=\ket{i}\bra{j}$, where $i,j\in \{\ket{g},\ket{e}\}$ are the transition operators of the TLS. The second term in the Hamiltonian accounts for the photon field at position $\bf{r}$ with the bosonic annihilation operators ${\bf\hat{f}({\bf r})}$. Finally, the last term accounts for the the light-matter interaction between the emitter dipole ${\bf \hat{d}}$ and the electric field ${\bf\hat{E}({\bf r})}={\bf\hat{E}^{+}({\bf r})}+{\bf\hat{E}^{-}({\bf r})}$. The response of the TLS can be expressed by the partially traced density matrix giving the elements $\rho_{ij}$.  In the rotating wave approximation and solving for the steady state  solution ($\dot{\hat{\rho}}=0$) we obtain the elements: 
\begin{equation}
\begin{split}
\rho_{ee}&= \frac{2\gamma_2 \Omega^2}{\gamma(\gamma_2^2+\Delta_2+4(\gamma_2/\gamma)\Omega^2)} \\
\rho_{ge}&= -\frac{ \Omega(i\gamma_2+\Delta)}{\gamma_2^2+\Delta^2+4(\gamma_2/\gamma)\Omega^2}
\end{split}
\label{EQ:DensityMatrix}
\end{equation}
Where $\gamma$ is the total emission rate that together with the pure dephasing rate $\gamma_{dp}$ constitutes $\gamma_2=\gamma/2+\gamma_{dp}$. % $\gamma_{deph}$, common in solid-state emitters, and modeled thanks to Lindblad operators. 
 While the population is also dependent on the driving field amplitude through Rabi frequency $\Omega={\bf d}\cdot {\bf E} /\hbar$.\\

In a single-mode conventional waveguide, the resulting transmitted "output" electric field can be expressed in terms of the input driving field\cite{Turschmann2019,kimble2017}:
\begin{equation}
{\bf\hat{E}^{+}_{out}({\bf r})}={\bf\hat{E}^{+}_{in}({\bf r})}+i\frac{\beta\gamma}{2\Omega}{\bf\hat{E}_{in}^{+}({\bf r})}\hat{\sigma}_{ge}
\label{EQ:ToalElectriField}
\end{equation}

Where waveguide-emitter coupling efficiency is governed by the ratio $\beta=\frac{\gamma_{WG}}{\gamma}$. Here $\gamma_{WG}$ is the rate of decay into the waveguide mode. The coupling factor is divided by 2 as equal coupling to both directions of propagation is assumed i.e the coupling is isotropic.
From this we define the corresponding transmission coefficient $t$ that transforms the input electric field $\bf\hat{E}^{+}_{in}({\bf r})$ to $\bf\hat{E}^{+}_{out}({\bf r})$ through the photonic waveguide. Using equation \ref{EQ:ToalElectriField}, results in:
\begin{equation}
    t=\frac{\langle{\bf\hat{E}_{out}^{+}({\bf r})}\rangle_{ss}}{\langle{\bf\hat{E}^{+}_{in}({\bf r})}\rangle_{ss}}=1+i\frac{\beta\gamma}{2\Omega}\rho_{eg}
\end{equation}

Inserting the density matrix element $\rho_{eg}=\rho^*_{ge}$ of equation (\ref{EQ:DensityMatrix}), we obtain:

\begin{equation}
    t=1-\frac{\beta\gamma}{2}\frac{ (\gamma_2+i\Delta)}{\gamma_2^2+\Delta^2+4(\gamma_2/\gamma)\Omega^2}
\end{equation}
Finally, the normalized intensity of the transmitted light can be calculated as:
\begin{equation}
\begin{split}
    I_t&=\frac{\langle{\bf\hat{E}^{-}_{out}({\bf r})}{\bf\hat{E}^{+}_{out}({\bf r})}\rangle_{ss}}{\langle{\bf\hat{E}^{-}_{in}({\bf r})}{\bf\hat{E}^{+}_{in}({\bf r})}\rangle_{ss}}\\
   &= 1-\frac{\beta\gamma\gamma_2(2-\beta)}{2(\gamma_2^2+\Delta^2+4(\gamma_2/\gamma)\Omega^2)}
    \end{split}
\end{equation}
We emphasize that $I_t\neq |t|^2$. 

 \subsubsection{Maximal phase shift}
The maxima of the phase shift with respect to the detuning can be found, at low power ($\Omega\ll	1$) and in the absence of dephasing, by solving:
\begin{equation}
    \frac{\partial \arg(t)}{\partial \Delta}(\Delta_{\pm})=\frac{2 \beta  \gamma  \left((\beta -1) \gamma ^2+4 \Delta_{\pm}^2\right)}{\left(\gamma ^2+4 \Delta_{\pm}^2\right) \left((\beta -1)^2 \gamma ^2+4 \Delta_{\pm}^2\right)}=0
\end{equation}
which corresponds to
\begin{equation}
    \Delta_{\pm}=\pm  \gamma  \frac{\sqrt{1-\beta}}{2}
\end{equation}
Plugging this back in the expression of the argument, one can find : 
\begin{equation}
\begin{split}
    |\phi|_{max} &=|\arg(t(\Delta_{\pm})|\\
    &=\arg \left(\frac{\left(2-i \sqrt{1-\beta }\right) \beta -2 }{\beta -2}\right)\\
    &=tan^{-1} \left( \frac{\beta }{2 \sqrt{1-\beta }}\right)\\
\end{split}
\end{equation}

\subsubsection{Transmission for a chirally coupled emitter}

In a waveguide with chiral light-matter coupling the interaction is directionally dependent. Similar to before, the total electric field in transmission is
\begin{equation}
{\bf\hat{E}^{+}_{out}({\bf r})}={\bf\hat{E}_{in}({\bf r})}+i\frac{\beta_{dir}\gamma}{\Omega}{\bf\hat{E}^{+}_{in}({\bf r})}\hat{\sigma}_{ge}
\end{equation}
where we define the directional coupling efficiency as $\beta_{dir}=\gamma_{t}/\gamma$, by differing the emission rate in transmitted($t$) or reflected modes($r$).
Following the same method as for conventional waveguide, we have:
\begin{equation}
\begin{split}
    &t_{dir}=1-\frac{\beta_{dir}\gamma (\gamma_2+i\Delta)}{\gamma_2^2+\Delta^2+4(\gamma_2/\gamma)\Omega^2}\\
    &I_{t_{dir}}=1+\frac{2\beta_{dir}\gamma\gamma_2(\beta_{dir}-1)}{\gamma_2^2+\Delta^2+4(\gamma_2/\gamma)\Omega^2}
        \end{split}
\end{equation}

Note that in the case of an isotropic, conventional waveguide  $(\gamma_{t}=\gamma_r=\gamma_{WG}/2)$, we recover the equation for an emitter coupled isotropically to waveguide modes.

\subsection{Mach-Zehnder interferometry}

\begin{figure}
  \includegraphics[width=\linewidth]{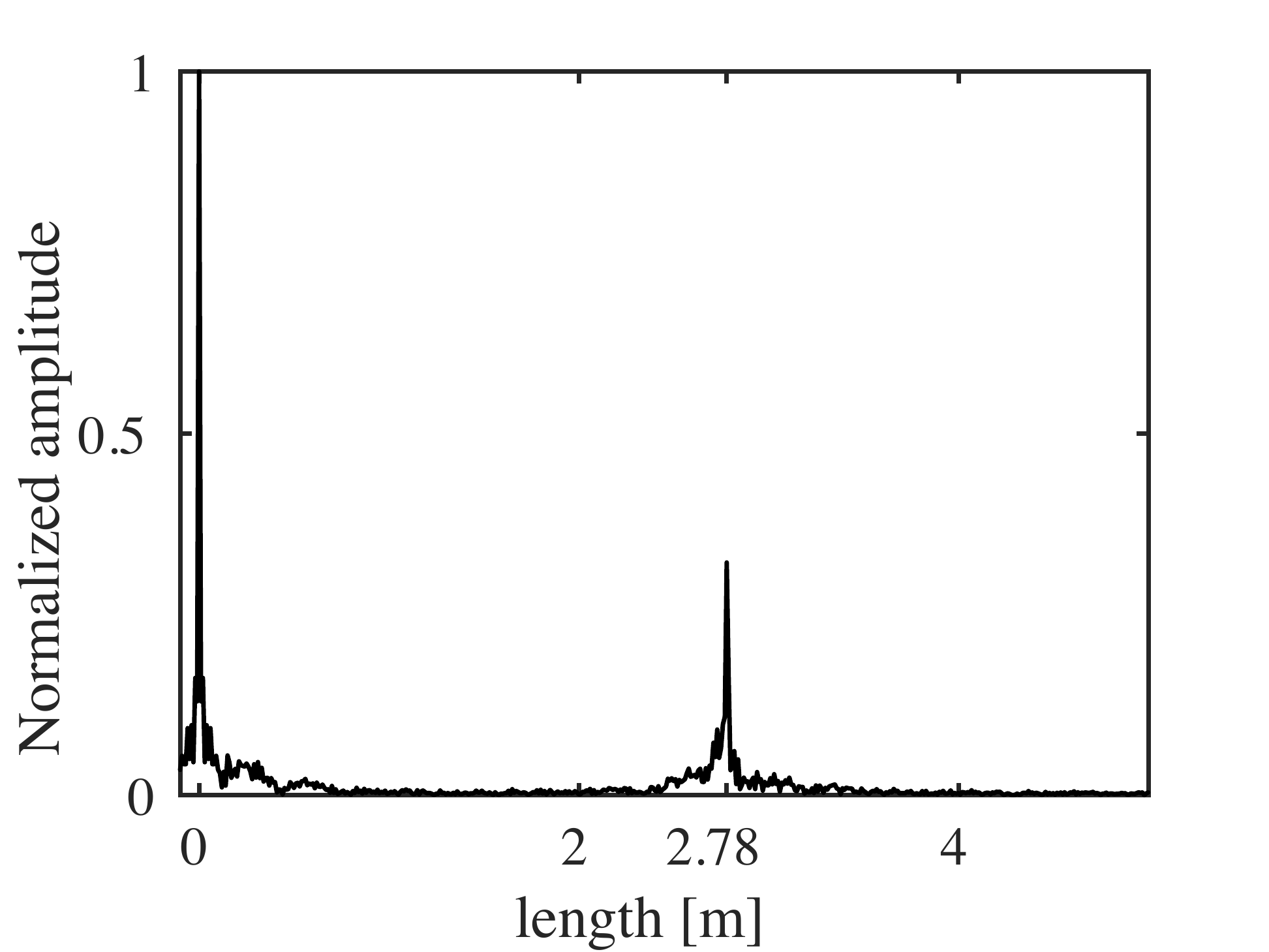}
  \caption{Normalized Fourier transformation of the interferometric signal as function of $\delta L$ in meters with the QD turned off.}
  \label{fig:SupFig1}
\end{figure}

The intensity of the output modes in a Mach-Zehnder interferometer is affected by the difference in phase, $\delta\phi$, between the two paths in the interferometer:
\begin{equation}
    I=\sin^2({\delta\phi/2})
    \label{EQ:Interferometer}
\end{equation}
 When light at frequency $f$ travels through each arm of the Mach-Zender interferometer (1,2), it experiences a phase shift of $\phi_{1,2}=2\pi f L_{1,2} / cn_{1,2}$, where the speed of light is c and the index of refraction $n$ may be different in the two arms with respective distances $L_{1,2}$. Additionally, there may be an environmental fluctuation phase difference $\delta\phi_{\text{env}}$.
Only one path (path 1) is affected by a phase change $\phi_{QD}=\text{arg}(t)$ induced by the quantum dot waveguide system. Therefore, the final interferometric phase difference can be expressed as
\begin{equation}
    \delta\phi = \phi_1 - \phi_2=  \frac{2\pi f \delta L }{c} + \delta\phi_{\text{env}} + \phi_{QD}
    \label{EQ:totdiffphase}
\end{equation}
$\delta L $ is the interferometric path length difference. 
The interferometric signal obtained when sweeping the laser detuning is displayed in Fig.~\ref{fig:figure1}(b).
The Fourier transform(FFT) of these interferometric fringes is displayed in Fig.\ref{fig:SupFig1} . Using Equations \ref{EQ:Interferometer} and \ref{EQ:totdiffphase}, we identify that the main frequency component of the fourier transform as $f=\delta L/c$ and we estimate the full path length difference of our interferometer to be $\delta L \approx 2.78 m$.

\section{Modeling the experimental data}
\subsection{Phase and Intensity}
\begin{figure}
\includegraphics[width=\linewidth]{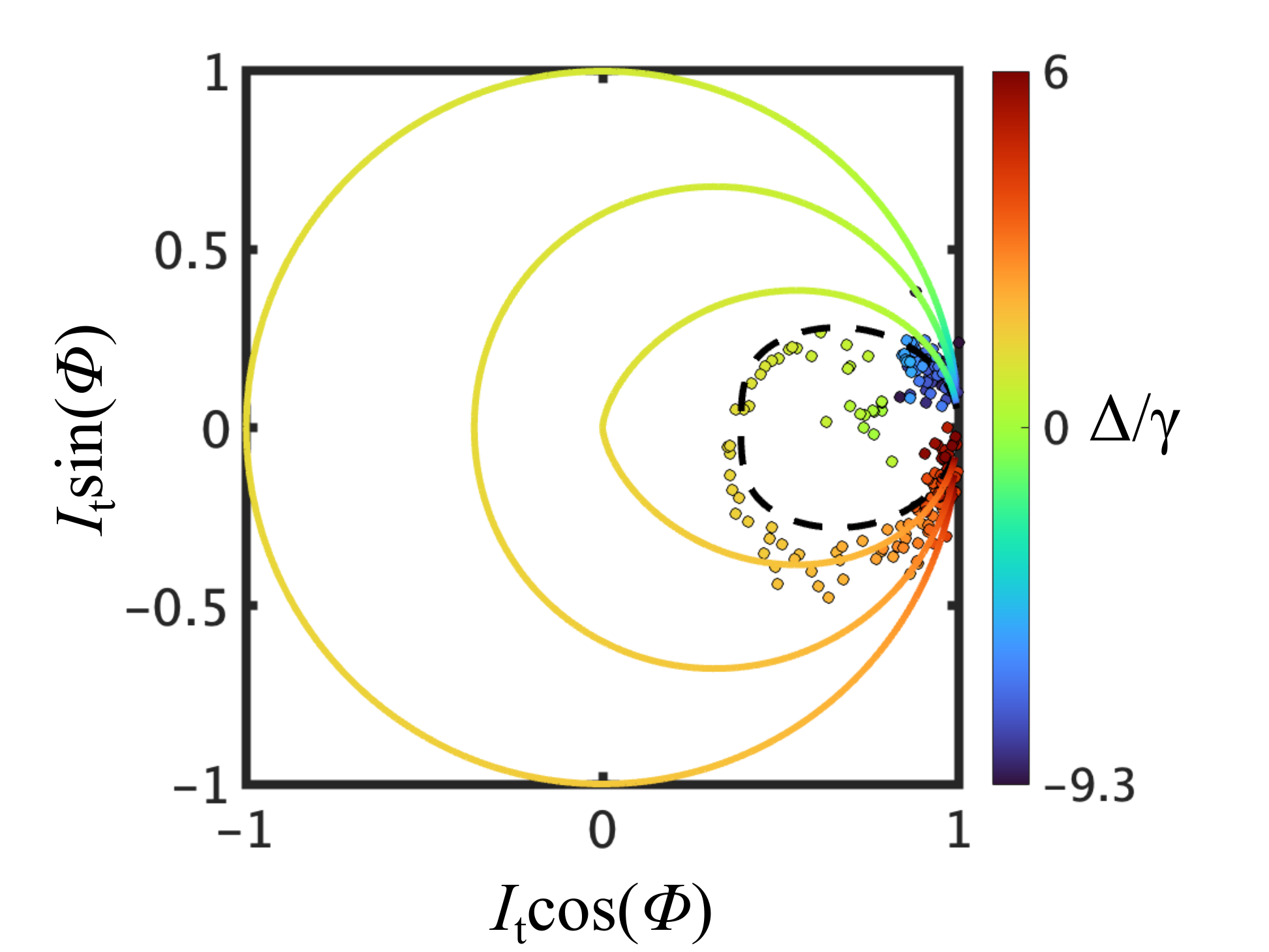}
    \centering
    \caption{Visualization of the effect of transition 2 (colored dots), and its corresponding fit (dashed black line) in a phasor diagram as a function of the normalized detunning $\Delta/\gamma$. For comparison, solid colored lines from inner to outer curves represent calculations for increasing directional coupling efficiencies $\beta_{dir}=\{0.5,0.8,1\}$.}
    \label{fig:SupFig2}
\end{figure}

We simultaneously fit the phase and intensity data of the two dipoles' response displayed in Fig.~\ref{fig:figure2}(c) and (d). We assume here for simplicity identical dephasing rates for both dipoles. Furthermore, we assume only pure dephasing, while in reality also slow noise processes (spectral diffusion) are influencing, however an unambiguous separation of these two processes is outside the scope of the present work; for more information, see \cite{LeJeannic2021}. As a consequence, the extracted pure dephasing rates will be overestimated. We adjusted the displayed data by taking into account the constant offset $\phi_0$ caused by weak Fano resonances, which are a result of partial reflection from the outcoupling gratings of the waveguide. (More information can be found in the references \cite{Javadi2015,LeJeannic2021})
We find the parameters to be:
\begin{table}[h!]
\centering
\begin{tabular}{ |c|c||c|}
 \hline
  &Dipole 1& Dipole 2 \\
 \hline
$\beta$ & $0.94\pm 0.03$& $ 1$\\
 \hline
  $\gamma~(\textrm{ns}^{-1})$ & $9.4\pm 0.2$    & $12.3\pm 0.2$.\\
 \hline
$\gamma_{dp}~(\textrm{ns}^{-1})$ & \multicolumn{2}{|c|}{$3.9 \pm 0.1$} \\
   \hline
$\phi_{0}$ (rad)& \multicolumn{2}{|c|}{ $-0.25 \pm 0.02$} \\
 \hline
\end{tabular}
\end{table}
\\
We present the data of dipole transition 2, and the corresponding model fit in a phasor diagram in Fig.~\ref{fig:SupFig2}

\subsection{Saturation Characterization}

In the following, we focus only on transition (2). We fit the saturation of the transmission spectra presented in Fig.~\ref{fig:figure3} with our model and obtain the following parameters:
\begin{table}[h!]
\centering
\begin{tabular}{ |c|c|}
 \hline
$\beta$ & $0.99~[0.57,1]$\\
 \hline
  $\gamma~(\textrm{ns}^{-1})$ & $12.6~[7.7, 17.4]$\\
 \hline
$\gamma_{dp}~(\textrm{ns}^{-1})$ & $3.4 [0,7.4]$ \\
   \hline
$\phi_0$ (rad) & $-0.26~ [-0.31,-0.2]$ \\
 \hline
\end{tabular}
\end{table}

Those values are in good agreement with the data of the two dipoles in Figure \ref{fig:figure2}.
We quantify the critical photon flux at the saturation as: $n_c=\frac{1+2\gamma_{dp}/\gamma}{4\beta^2}\sim0.33$ \cite{Javadi2015,LeJeannic2021}.
At each power value we also perform an independent fit with free parameters, to extract accurately the maximal, experimentally measured phase shift from the data. Those are the data points $|\phi_{max}|$ displayed in Fig.~\ref{fig:figure3}(b).
%The referenced data samples of (a) and (b) on \ref{fig:figure2} correspond respectively to phase shifts of $\phi_a=-0.18\pm0.09$ and $\phi_b=-0.59\pm0.08$ radians.

\end{document}